\documentclass[aps,pra,reprint,twocolumn,nofootinbib,a4paper,floatfix,10pt]{revtex4-1}

\usepackage{color}
\usepackage{bm}
\usepackage{amsmath,amssymb}
\usepackage{amsfonts}
\usepackage{dsfont}
\usepackage{graphicx}
\usepackage{float}
\usepackage{subfigure}
\usepackage{verbatim}
\usepackage{amsfonts}
\usepackage{dcolumn}
\usepackage{bm}
\usepackage{color}
\usepackage[dvipsnames]{xcolor}
\usepackage{listings}
\usepackage{mathrsfs}
\usepackage{filecontents}
\usepackage{framed}
\usepackage{times} %font times new roman

\definecolor{blueref}{rgb}{0.2, 0.3, 0.7}
\definecolor{bluemathematica}{rgb}{0.368417, 0.506779, 0.709798}
\definecolor{yellowmathematica}{rgb}{0.880722, 0.611041, 0.142051}
\usepackage[colorlinks=true,urlcolor=blueref,citecolor=blueref,linkcolor=blueref]{hyperref}
\usepackage{mathtools}

\lstnewenvironment{mat}
{\lstset{language=mathematica,mathescape,columns=flexible,basicstyle=\ttfamily\footnotesize}}
{}

\newcommand{\bra}[1]{\mbox{$\langle #1 |$}}
\newcommand{\ket}[1]{\mbox{$| #1 \rangle$}}
\newcommand{\tr}{\mbox{tr}}

\definecolor{blue}{rgb}{0.368417,0.506779,0.709798}

\begin{document}

\title{Quantifying entanglement of formation for two-mode Gaussian states: Analytical expressions for upper and lower bounds and numerical estimation of its exact value}
\author{Spyros Tserkis}  \email{s.tserkis@uq.edu.au}
\affiliation{Centre for Quantum Computation and Communication Technology, School of Mathematics and Physics, University of Queensland, St Lucia, Queensland 4072, Australia}
\author{Sho Onoe} 
\affiliation{Centre for Quantum Computation and Communication Technology, School of Mathematics and Physics, University of Queensland, St Lucia, Queensland 4072, Australia}
\author{Timothy C. Ralph} 
\affiliation{Centre for Quantum Computation and Communication Technology, School of Mathematics and Physics, University of Queensland, St Lucia, Queensland 4072, Australia}

\begin{abstract}
Entanglement of formation quantifies the entanglement of a state in terms of the entropy of entanglement of the least entangled pure state needed to prepare it. An analytical expression for this measure exists only for special cases, and finding a closed formula for an arbitrary state still remains an open problem. In this work we focus on two-mode Gaussian states, and we derive narrow upper and lower bounds for the measure that get tight for several special cases. Further, we show that the problem of calculating the actual value of the entanglement of formation for arbitrary two-mode Gaussian states reduces to a trivial single parameter optimization process, and we provide an efficient algorithm for the numerical calculation of the measure.
\end{abstract}

\maketitle

\section{Introduction}
\label{sec1}

Quantifying entanglement is a non-trivial task, since various measures exist with different operational meanings, and most of them lack an analytical expression. Every entanglement measure $\mathcal{E}$ needs to satisfy the following postulates \cite{Horodecki.et.al.RMP.09,Plenio.Virmani.B.14}:  (i) $\mathcal{E}$ vanishes on separable states, (ii) $\mathcal{E}$ does not increase on average under local operations and classical communication (strong monotonicity), and, (iii) for pure states $\mathcal{E}$ is equal to the entropy of entanglement, given by the von Neumann entropy of the reduced state.

Among several entanglement measures, entanglement of formation (EoF) is of significant importance, due to its well-defined physical meaning, i.e., EoF quantifies the entanglement of a state in terms of the entropy of entanglement of the least entangled pure state needed to prepare it \cite{Bennett.DiVincenzo.et.al.PRA.96}. For a given state $\hat{\sigma}:=\sum_i p_i \ket{\psi_i}\bra{\psi_i}$, EoF is given by the convex-roof extension of the reduced von Neumann entropy of $\ket{\psi_i}$, i.e.,
\begin{equation}
\mathcal{E}_F(\hat{\sigma}):=\inf_{\{p_i,\psi_i\}} \{\sum_i p_i \mathcal{H}(\tr_B\ket{\psi_i}\bra{\psi_i})\} \,.
\label{EoFgeneral}
\end{equation}

In general the calculation of EoF is NP-hard (non-deterministic polynomial-time hard) \cite{Huang.NJP.14} and there are only few cases, e.g., for qubits \cite{Wootters.PRL.98}, where Eq.~(\ref{EoFgeneral}) reduces to an analytical expression.
 
In this paper we work with systems of quantized radiation modes of the electromagnetic field that are described by continuous-variable states  \cite{Serafini.B.17,Weedbrook.et.al.RVP.12,Adesso.Ragy.OSID.14}. Those modes are associated with the quadrature field operators $\hat{x}_j:=\hat{a}_j+\hat{a}^{\dag}_j$ and $\hat{p}_j:=i(\hat{a}^{\dag}_j-\hat{a}_j)$, where $\hat{a}_j$ and $\hat{a}^{\dag}_j$ are the annihilation and creation operators, respectively, with $[\hat{a}_i,\hat{a}^{\dag}_j]=\delta_{ij}$. We specifically focus on two-mode Gaussian states, which can be fully described by the first two statistical moments of the quadratures field operators.

In particular, we derive an upper bound for the entanglement of formation that comes as an extension of our recently derived lower bound for two-mode Gaussian states \cite{Tserkis.Ralph.PRA.17}. We also present a new optimization method for estimating the real value of the measure, supplemented by an explicit algorithm written in \textit{Mathematica} for the numerical estimation of the measure \cite{supplementalmaterial}. A numerical comparison of the lower and upper bounds to the exact value of the EoF is also presented for a set of randomly created states against their global purity.

In Sec.~\ref{sec2} we briefly review the structure of two-mode Gaussian states along with their classicality and separability conditions. In Sec.~\ref{sec3} we start by defining entanglement of formation, for the general case, and we continue by presenting the lower bound derived in Ref.~\cite{Tserkis.Ralph.PRA.17} in order to use it for the derivation of the upper bound. We also introduce a new simple optimization method for the estimation of the real value of the measure for arbitrary states. Finally, we see how close the upper and lower bounds are to the actual EoF for randomly created entangled states. In Sec.~\ref{sec4} we conclude our work.

\section{Gaussian states}
\label{sec2}

\subsection{State representation}

A two-mode Gaussian state $\hat{\sigma}$ with zero mean value (for simplicity) can be fully described by its covariance matrix
\begin{equation}
\boldsymbol{\sigma} := \begin{bmatrix}
\boldsymbol{A} & \boldsymbol{C} \\
\boldsymbol{C}^T & \boldsymbol{B} 
\end{bmatrix}\,,
\label{cov}
\end{equation}
which is a real, symmetric, and positive definite matrix with elements proportional to the second-order moments of the quadrature field operators. The global purity of the state is given by $\mu:=1/\sqrt{\det \boldsymbol{\sigma}}$, while local purities by $\mu_a:=1/\sqrt{\det \boldsymbol{\boldsymbol{A}}}$ and $\mu_b:=1/\sqrt{\det \boldsymbol{\boldsymbol{B}}}$, respectively. In the standard form \cite{Duan.et.al.PRL.00,Simon.PRL.00}, the covariance matrix $\boldsymbol{\sigma}^{\text{sf}}$ is given by $\boldsymbol{A}=\text{diag}(a,a)$, $\boldsymbol{B}=\text{diag}(b,b)$, with $a \geqslant b$, and $\boldsymbol{C}=\text{diag}(c_1,c_2)$, with $c_1 \geqslant |c_2| \geqslant 0$. The elements of the covariance matrix in the standard form can be parametrized over the local and global purities of the state as follows \cite{Adesso.Serafini.Illuminati.PRA.04}
\begin{align}
a &= \frac{1}{\mu_a}\,,  \quad \quad c_1 = \frac{z+w}{8} \sqrt{\mu_a \mu_b} \,,\\
b &= \frac{1}{\mu_b}\,, \quad \quad c_2 = \frac{z-w}{8}  \sqrt{\mu_a \mu_b} \,,
\end{align}
where
\begin{align}
z &= \sqrt{\left[8d^2{+}(\beta {-}1) (1{+}g^2){-}2(\beta{+}1) (2d^2{+}g)\right]^2{-}16 g^2} \,, \\
w &= \sqrt{\left[8s^2{+}(\beta {-}1) (1{+}g^2){-}2(\beta{+}1) (2d^2{+}g)\right]^2{-}16 g^2} \,,
\end{align}
with $s=(a+b)/2$, $d=(a-b)/2$, $g=1/\mu$, and $-1 \leqslant \beta \leqslant 1$ (In Refs.~\cite{Adesso.Serafini.Illuminati.PRA.04,Adesso.Illuminati.PRA.05} states with $\beta=1$ are called GMEMS and states with $\beta=-1$ are called GLEMS). The parameters $s$, $d$, and $g$ are constrained as follows: $s \geqslant 1$, $|d| \leqslant s-1$, and $g \geqslant 2 |d|+1$.

\subsection{Classicality}

Every quantum state $\hat{\sigma}$ can be represented in phase-space with the so-called $\mathcal{P}$-function \cite{Glauber.PR.63,Sudarshan.PRL.63} defined as
\begin{equation}
\hat{\sigma}:=\int \mathcal{P}(\alpha) \ket{\alpha} \bra{\alpha} d^2 \alpha \,,
\end{equation}
where $\ket{\alpha}$ represents a coherent state and $\mathcal{P}(\alpha)$ is a quasi-probability distribution. When the $\mathcal{P}$-function takes positive values it can be interpreted as a classical probability distribution and the corresponding state is called classical, and when the $\mathcal{P}$-function is negative or singular the corresponding state is called non-classical \cite{Mandel.PS.86}.

For two-mode Gaussian systems a state is classical if and only if $\boldsymbol{\sigma}^{\text{sf}} \geqslant \mathds{1}_4$, i.e., all the eigenvalues of its covariance matrix are greater or equal to 1 \cite{Serafini.B.17}. 

\subsection{State decomposition}

According to Williamson's theorem \cite{Williamson.AJM.36}, for every covariance matrix there is a symplectic transformation $\boldsymbol{K}$, such that 
\begin{equation}
\boldsymbol{\sigma} = \boldsymbol{K} [\nu_-\mathds{1}_2 \oplus \nu_+\mathds{1}_2]\boldsymbol{K}^{T} \,,
\label{Williamson}
\end{equation}
with  $1 \leqslant \nu_- \leqslant \nu_+$ being the symplectic eigenvalues \cite{Vidal.Werner.PRA.02}, given by \cite{Serafini.Illuminati.DeSiena.JPB.04}
\begin{equation}
\nu_{\pm}=\sqrt{\frac{\Delta \pm \sqrt{\Delta^2-4 \det \boldsymbol{\sigma} }}{2}}  \,,
\end{equation}
where $\Delta=\det \boldsymbol{A} +\det \boldsymbol{B} +2 \det \boldsymbol{C}=\nu_-^2+\nu_+^2 \geqslant 1$ is invariant under global symplectic operations. Re-arranging Eq.~(\ref{Williamson}) we get
\begin{equation}
\boldsymbol{\sigma} = \boldsymbol{\sigma}_p + \boldsymbol{\phi} \,,
\label{dec}
\end{equation}
where $\boldsymbol{\sigma}_p$ is a pure state, also called two-mode squeezed vacuum, that in the standard form is given by
\begin{equation}
\boldsymbol{\sigma}_p^{\text{sf}} := \begin{bmatrix}
\cosh (2r) \, \mathds{1}_2 & \sinh (2r) \, \boldsymbol{Z} \\
\sinh (2r) \, \boldsymbol{Z} & \cosh (2r) \, \mathds{1}_2
\end{bmatrix}\,,
\end{equation}
with $r \in \mathbb{R}$, where $\boldsymbol{Z}=\text{diag(1,-1)}$, and $\boldsymbol{\phi} \geqslant \boldsymbol{0}$ is a positive semi-definite matrix. Equivalently, Eq.~(\ref{Williamson}) can also be written as a symplectic transformation $\boldsymbol{\Sigma}$ applied on a classical state $\boldsymbol{\sigma}_{\text{c}}$, i.e.,
\begin{equation}
\boldsymbol{\sigma} = \boldsymbol{\Sigma} \boldsymbol{\sigma}_{\text{c}} \boldsymbol{\Sigma}^{T} \,.
\end{equation}

\begin{figure}[t]
\centering
  \includegraphics[width=\columnwidth]{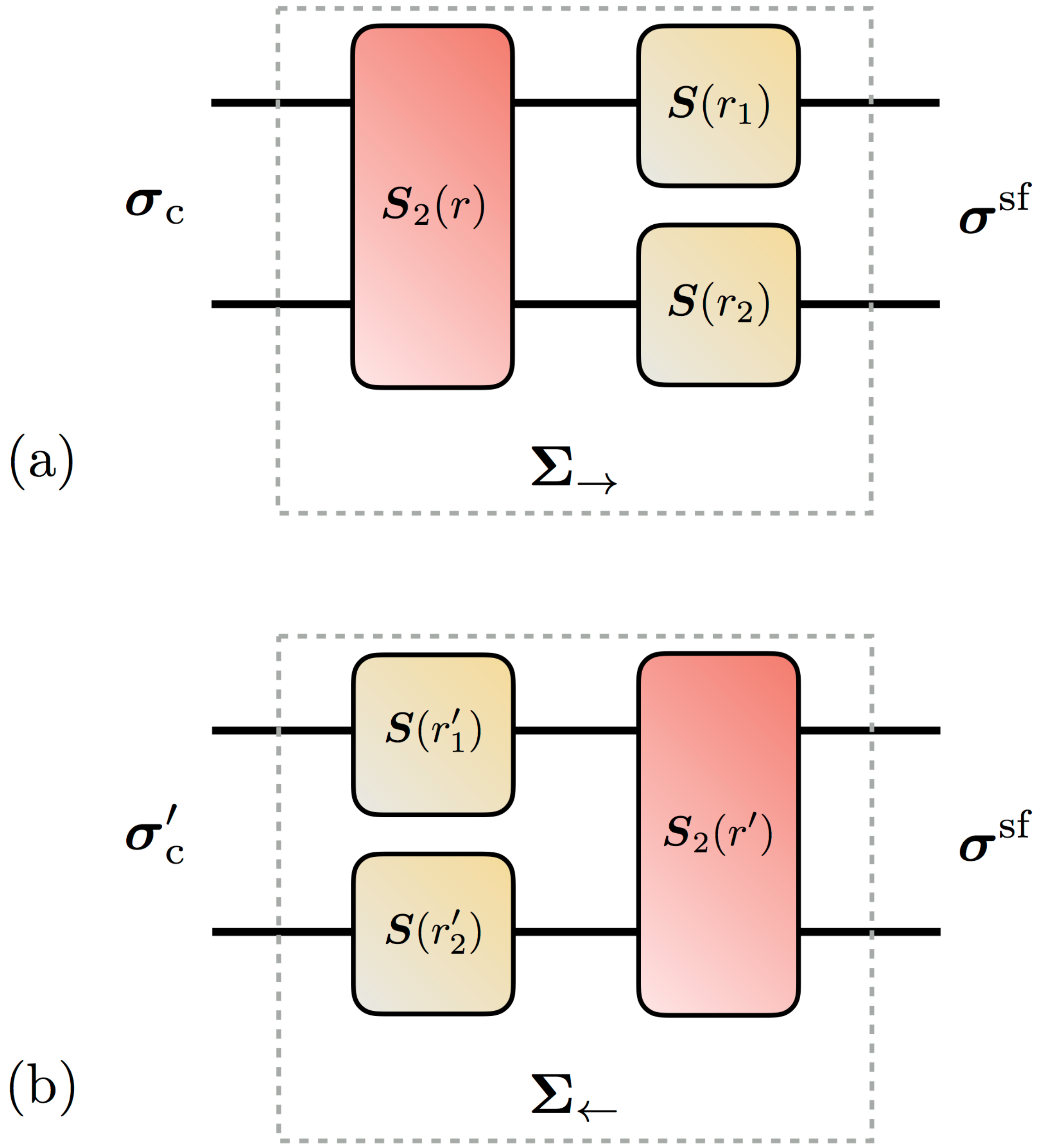}
  \caption{ \small In figure (a) and (b) we present two symplectic transformations $\boldsymbol{\Sigma}_{\rightarrow}$ and $\boldsymbol{\Sigma}_{\leftarrow}$, given in Eq.~(\ref{transf1}) and Eq.~(\ref{transf2}), respectively. Both of them are decomposed into a sequence (direct and reverse) of a two-mode squeezing transformation $\boldsymbol{S}_2$ and two single-mode squeezing transformation $\boldsymbol{S}$. Every state in the standard form can prepared by applying $\boldsymbol{\Sigma}_{\rightarrow}$ or $\boldsymbol{\Sigma}_{\leftarrow}$ onto a classical state.}
  \label{fig1}
\end{figure}

Two decompositions relevant to our following analysis (graphically presented in Fig.~\ref{fig1}) are the following
\begin{equation}
\boldsymbol{\Sigma}_{\rightarrow} := \boldsymbol{L}(r_1,r_2) \boldsymbol{S}_2(r)\,,
\label{transf1}
\end{equation}
and its transpose, i.e.,
\begin{equation}
\boldsymbol{\Sigma}_{\leftarrow} := \boldsymbol{S}_2(r') \boldsymbol{L}(r'_1,r'_2)\,,
\label{transf2}
\end{equation}
with $\boldsymbol{L}(r_1,r_2):=\boldsymbol{S}(r_1)\oplus \boldsymbol{S}(r_2)$, where $\boldsymbol{S}(r_i):=\exp[r_i \boldsymbol{Z}]$ is the local squeezing symplectic operation, i.e.,
\begin{equation}
\boldsymbol{L}(r_1,r_2) := \exp \begin{bmatrix}
r_1 \, \boldsymbol{Z} & 0 \\
0 & r_2 \, \boldsymbol{Z}
\end{bmatrix} \,,
\end{equation}
and $\boldsymbol{S}_2(r)$ is the two-mode squeezing symplectic operation given by 
\begin{equation}
\boldsymbol{S}_2(r) := \begin{bmatrix}
\cosh r \, \mathds{1}_2 & \sinh r \, \boldsymbol{Z} \\
\sinh r \, \boldsymbol{Z} & \cosh r \, \mathds{1}_2
\end{bmatrix}\,.
\end{equation}

\subsection{Separability}

Witnessing entanglement for arbitrary states is in general a difficult problem, however, in two-mode Gaussian states the separability criterion, also called the Peres-Horodecki criterion \cite{Peres.PRL.96,Horodecki.PLA.96}, is necessary and sufficient \cite{Duan.et.al.PRL.00,Werner.Wolf.PRL.01,Simon.PRL.00}. In particular, the separability of such states can be checked by the lowest symplectic eigenvalue of the partially transposed covariance matrix $\boldsymbol{\sigma}^{\Gamma}=(\mathds{1} \oplus \boldsymbol{Z}) \boldsymbol{\sigma} (\mathds{1} \oplus \boldsymbol{Z})$, i.e., separable states are the ones with $\nu^{\Gamma}_- \geqslant 1$ \cite{Adesso.Serafini.Illuminati.PRA.04}, where 
\begin{equation}
\nu_{\pm}^{\Gamma}=\sqrt{\frac{E \pm \sqrt{E^2-4 \det \boldsymbol{\sigma}}}{2}}  \,,
\end{equation}
with $E=\det \boldsymbol{A} +\det \boldsymbol{B} - 2 \det \boldsymbol{C}=\left( \nu_-^{\Gamma} \right)^2+\left( \nu_+^{\Gamma} \right)^2 \geqslant 1$.

\section{Entanglement of Formation}
\label{sec3}

Entanglement of formation (EoF) for a two-mode Gaussian state $\boldsymbol{\sigma}$ coincides with the Gaussian entanglement of formation (GEoF) \cite{Akbari-Kourbolagh.Alijanzadeh-Boura.QIP.15} and is equal to \cite{Wolf.et.al.PRA.04,Ivan.Simon.arXiv.08,Marian.Marian.PRL.08}
\begin{equation}
\mathcal{E}_F (\boldsymbol{\sigma}) := \inf_{{\boldsymbol{\sigma}_p}_i} \{ \mathcal{H}[{\boldsymbol{\sigma}_p}_i (r) ] \, | \, \boldsymbol{\sigma}={\boldsymbol{\sigma}_p}_i+\boldsymbol{\phi}_i \} \,,
\label{eof}
\end{equation}
where $\mathcal{H}$ is the entropy of entanglement of a pure state $\boldsymbol{\sigma}_p$ with a two-mode squeezing parameter $r$, i.e., \cite{Holevo.Sohma.Hirota.PRA.99}
\begin{equation}
\mathcal{H}[\boldsymbol{\sigma}_p (r) ] := \cosh^2 r \log_2 ( \cosh^2 r)-\sinh^2 r \log_2 ( \sinh^2 r) \,.
\label{entropyofentanglement}
\end{equation}

The optimal decomposition corresponds to the pure state (two-mode squeezed vacuum) with the least entropy of entanglement that can be transformed under local operations and classical communication into our state. From a resource theoretic point of view, the optimum decomposition corresponds to the minimum amount of two-mode squeezing needed for the creation of this pure state \cite{Tserkis.Ralph.PRA.17}.

The first attempt to derive a closed formula of this measure for mixed states was done by Giedke et.al. \cite{Giedke.et.al.PRL.03} in 2003, who gave an analytical expression of EoF for all symmetric states, i.e., $a=b$. Two years later, Adesso et.al. \cite{Adesso.Illuminati.PRA.05} managed to give an analytical formula for GEoF (which was later shown to be equivalent with EoF) for all mixed states with $\nu_-=1$, called GLEMS ($\beta=-1$), and for states with $c_1=-c_2$, also called GMEMS ($\beta=1$). In order to calculate numerically the exact value of the measure, we can follow the approaches of either Wolf et.al. \cite{Wolf.et.al.PRA.04}, Marians \cite{Marian.Marian.PRL.08} or Ivan and Simon \cite{Ivan.Simon.arXiv.08}. Later in Sec.~\ref{sec3} we will show how we can simplify the numerical calculation and calculate the EoF with a trivial optimization over a single parameter. 

Analytical lower bounds of the EoF have also been derived in Refs.~\cite{Rigolin.Escobar.PRA.04} and \cite{Nicacio.Oliveira.PRA.14}. Finally, in 2017, a narrow lower bound was derived \cite{Tserkis.Ralph.PRA.17}, that is consistently closer to the actual value of the measure compared to the older ones, and has also the advantage of being tight for symmetric and states with $\beta=1$.

\subsection{Lower Bound for Entanglement of Formation}

In Ref.~\cite{Tserkis.Ralph.PRA.17} we derived a lower bound for the entanglement of formation. In this section we re-derive it in a more elegant and compact way.

Let us assume all the possible decompositions of a state
\begin{equation}
\boldsymbol{\sigma}={\boldsymbol{\sigma}_p}_i+\boldsymbol{\phi}_i \,.
\label{decomposition}
\end{equation}

Among all pure states ${\boldsymbol{\sigma}_p}_i$ that satisfy the above decomposition, one has the minimum entropy of entanglement, i.e., the optimum pure state ${\boldsymbol{\sigma}_p}_o$. Using this optimal state, we are able to calculate the EoF of the state $\boldsymbol{\sigma}$ as follows
\begin{equation}
\mathcal{E}_F(\boldsymbol{\sigma}) =\mathcal{H}(r_o)  \,.
\end{equation}
 
Two (but not the only) ways to construct a pure state $\boldsymbol{\sigma}_{p_i}$ is by applying the symplectic transformations $\Sigma_{\rightarrow}$ or $\Sigma_{\leftarrow}$ onto a couple of vacua. For every two-mode squeezing parameter $r_i$ of the transformation $\Sigma_{\rightarrow}$ there is a corresponding parameter $r'_i$ of the transformation $\Sigma_{\leftarrow}$. It is easy to show that $r'_i \leqslant r_i$ for any pure state ${\boldsymbol{\sigma}_p}_i$ (this can be easily seen from Eq.~(\ref{upperr}) since value $r'$ is the global minimum of $r$). Thus the global minimums of the two-mode squeezing parameters $r_i$ and $r'_i$ of those two decompositions have the following ordering
\begin{equation}
\min_i \{ r'_i \} \equiv r_-  \leqslant  r_o \equiv \min_i \{ r_i \}  \,.
\end{equation}

The above equation essentially implies that the least amount of two-mode squeezing we need to apply to a state to make it separable is always less or equal to the least amount of two-mode squeezing we need to create it.

Assuming a state in the standard form $\boldsymbol{\sigma}^{\text{sf}}$, the lowest value of the two-mode squeezing $r'$ corresponding to the symplectic transformation $\Sigma_{\leftarrow}$ has been calculated in Ref.~\cite{Tserkis.Ralph.PRA.17}, and is equal to
\begin{equation}
r_-=\frac{1}{2} \ln \sqrt{\frac{\kappa - \sqrt{\kappa^2 - \lambda_+ \lambda_-}}{\lambda_-}} \,,
\label{lowerr}
\end{equation}
where we have set  $\kappa{= }2(\det \boldsymbol{\sigma} {+} 1){-}(a{-}b)^2$ and $\lambda_{\pm}{=}\det \boldsymbol{A}{+} \det \boldsymbol{B}{-}2\det \boldsymbol{C}{+}2[(a b {-} c_1 c_2){\pm} (c_1{-}c_2)(a{+}b)]$. Thus, for entangled states ($\nu^{\Gamma}_-<1$) substituting $r_-$ into the monotonic function given in Eq.~(\ref{entropyofentanglement}) we get a lower bound for the EoF, i.e.,
\begin{equation}
\nu^{\Gamma}_-(\boldsymbol{\sigma}^{\text{sf}})<1 \, \Rightarrow \,  \mathcal{E}_F^- (\boldsymbol{\sigma}^{\text{sf}}) = \mathcal{H}(r_-) \leqslant \mathcal{E}_F(\boldsymbol{\sigma}^{\text{sf}}) \,. 
\end{equation}

This lower bound is in general quite close to the actual value (see Fig.~\ref{fig2}) and it becomes tight when the transformation $\Sigma_{\rightarrow}$ that corresponds to the optimal decomposition of the state $\boldsymbol{\sigma}^{\text{sf}} $ is equivalent to the $\Sigma_{\leftarrow}$. It is trivial to show that $[\boldsymbol{S}_2(r), \boldsymbol{L}(r_{\ell},r_{\ell})]=0$, which means that when the two single-mode squeezers of either transformation $\Sigma_{\leftarrow}$ or $\Sigma_{\rightarrow}$ are equal to each other, they can commute through the two-mode squeezer and thus $\Sigma_{\leftarrow} \equiv \Sigma_{\rightarrow}$. That is true for both symmetric states and states with $\beta=1$ \cite{Tserkis.Ralph.PRA.17}.

It is also worth mentioning that the single-mode squeezing parameters $r'_1$ and $r'_2$ can also be analytically calculated for a given value of two-mode squeezing $r'$, i.e.,
\begin{align}
r'_1&= \ln \sqrt{\frac{(a-b) \xi_+ - 2 \theta \sinh (2 r')  - (a+b) \xi_- \cosh (2 r')}{\omega -\det \boldsymbol{\sigma} +1 +\sqrt{\gamma (\zeta_1 + \zeta_2)} }} \,, \\
r'_2&=\ln \sqrt{\frac{(a-b) \xi_+ + 2 \theta \sinh (2 r')  + (a+b) \xi_- \cosh (2 r')}{\omega +\det \boldsymbol{\sigma} - 1 +\sqrt{\gamma (\zeta_1 + \zeta_2)} }} \,.
\end{align}
with 
\begin{gather}
\xi_{\pm}= a b{-}c_1^2\pm1 \,, \\
\theta= a b c_2 {-}c_1^2 c_2{+}c_1\,, \\
\omega= (a{-}b) [(a+b) \cosh (2 r'){+}(c_2{-}c_1) \sinh (2 r')] \,, \\
\gamma= \frac{1}{2} [a^2 (b^2{-}1){-}a b (c_1^2{+}c_2^2){-}b^2{+}(c_1 c_2{-}1)^2 ] \,, \\
\zeta_1 = a^2 (2 b^2{-}1){-}2 a b \left(c_1^2{+}c_2^2{-}1\right){-}b^2{+}2 c_1^2 c_2^2{+}2 \,, \\
\zeta_2 = 2 (a{+}b) (c_1{-}c_2) \sinh (4 r'){-}\cosh (4 r') [ (a{+}b)^2{-}4 c_1 c_2 ] \,.
\end{gather}

\subsection{Upper Bound for Entanglement of Formation}

By definition of the measure, the entropy of entanglement of every pure state that satisfies the decomposition of Eq.~(\ref{decomposition}) constitutes an upper bound to the EoF. Every pure state created by the symplectic decomposition $\Sigma_{\leftarrow}$ applied onto a couple of vacua, can also be created by the symplectic decomposition $\Sigma_{\rightarrow}$ applied onto a couple of vacua. 

Let us use as a reference the pure state prepared with the transformation $\Sigma_{\leftarrow}$, with two-mode squeezing $r'$, and single-mode squeezing parameters $r'_1$ and $r'_2$. The equivalent pure state prepared with the transformation $\Sigma_{\rightarrow}$ has two-mode squeezing equal to
\begin{equation}
k(r')=\frac{1}{2} \cosh ^{-1}\left[e^{2 r'_2} \chi \sinh ^2 r' + e^{2 r'_1} \chi \cosh ^2 r' \right] \,,
\label{upperr}
\end{equation}
with 
\begin{equation}
\chi=\sqrt{\frac{e^{-2 r'_1}+e^{-2 r'_2} \tanh ^2 r'}{e^{2 r'_1}+e^{2 r'_2} \tanh ^2 r'}} \,.
\end{equation}
Setting $r'=r_-$ for entangled states ($\nu^{\Gamma}_-<1$), and by substituting this value into Eq.~(\ref{entropyofentanglement}) we get an upper bound for the entanglement of formation 
\begin{equation}
\nu^{\Gamma}_-(\boldsymbol{\sigma}^{\text{sf}})<1 \, \Rightarrow \, \mathcal{E}_F^+ (\boldsymbol{\sigma}^{\text{sf}}) = \mathcal{H}[k(r_-)]  \geqslant \mathcal{E}_F(\boldsymbol{\sigma}^{\text{sf}})\,,
\label{upperbound}
\end{equation}
that is actually quite narrow to the real value (see Fig.~\ref{fig2}). It is apparent that based on the way the upper and lower bound are connected, when the lower one gets tight the upper one gets tight as well (which happens for symmetric states and states with $\beta= 1$). After numerical calculations it seems that the upper bound becomes tight also for the case of states with $\beta= -1$, if the condition $ |r'_1-r'_2| \leqslant \frac{1}{2} \ln \nu_+$ is satisfied, but the general validity of this argument is only conjectured.

\subsection{Estimating Entanglement of Formation}

Finding an analytical expression for the exact value of the entanglement of formation is still considered an open problem. In this section we re-define EoF through a straightforward optimization process that involves the minimization over a single parameter.

As we discussed in the previous section, Eq.~(\ref{upperbound}) is in general an upper bound for EoF, since any valid pure state of Eq.~(\ref{decomposition}) has entropy of entanglement equal or greater to the optimal one. In order to find the optimal one we could minimize the upper bound over every possible pure state, however since we already have the squeezing values for the lower and upper bound we can express EoF as
\begin{align}
\mathcal{E}_F(\boldsymbol{\sigma}^{\text{sf}}) := \inf_{r'} \left\{ \mathcal{H}[k(r')] \, | \, r_- \leqslant r' \leqslant r_+\right\} \,.
\label{eof2}
\end{align}
The problem of writing down Eq.~(\ref{eof2}) as a closed formula is that the function that needs to be optimized is in general non-smooth. As mentioned before, though, for the cases of symmetric states, i.e., $a=b$, and states with $\beta= 1$, we don't need to optimize Eq.~(\ref{eof2}), since we just have to set $r'=r_-=r_+$. 

\begin{figure}[t]
\centering
  \includegraphics[width=\columnwidth]{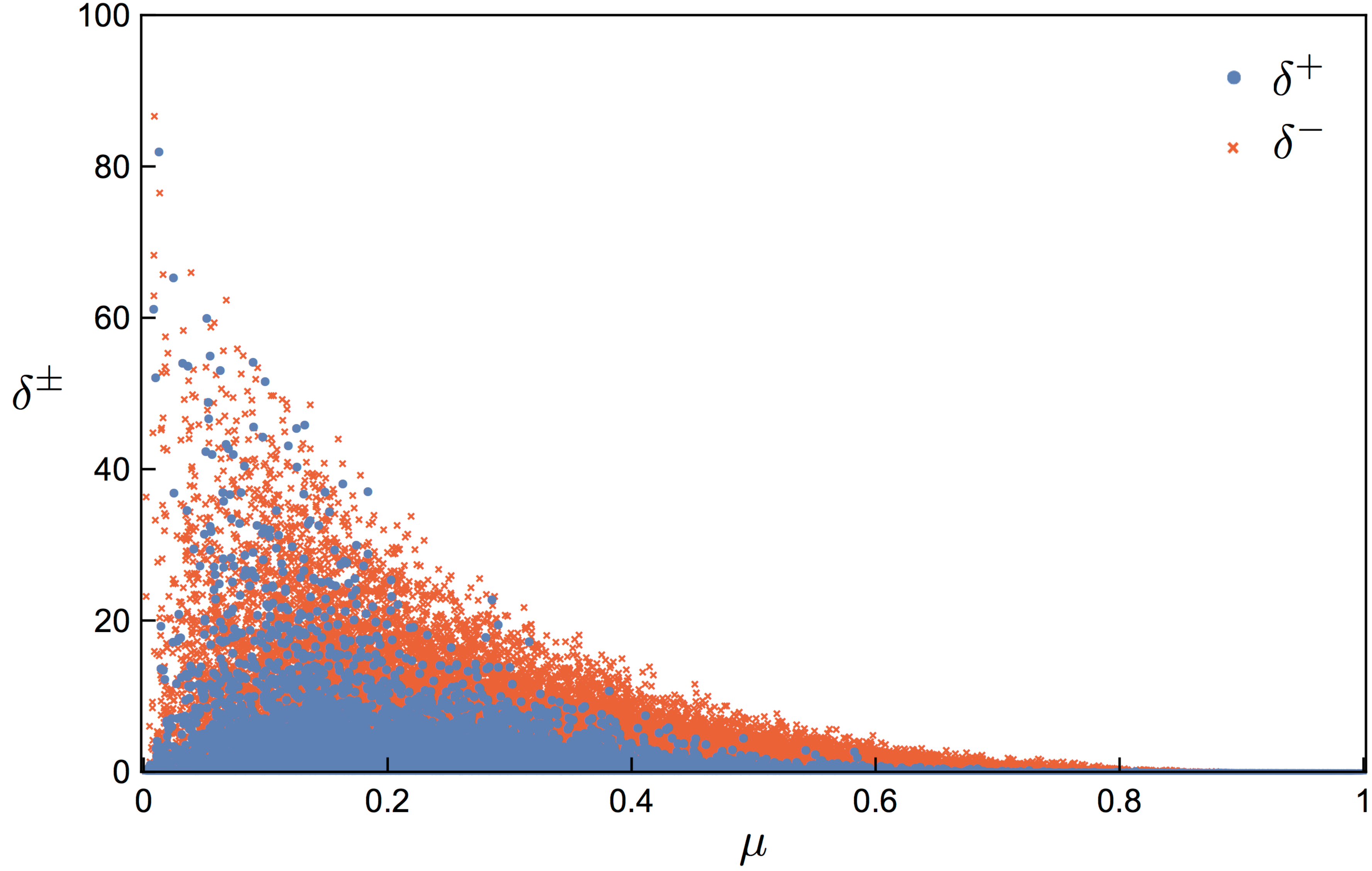}
  \caption{ \small In this figure we plot the percentile relative difference between both the upper $\delta^{+}$ (blue dots) and lower $\delta^{-}$ (red crosses) bound and the actual value of the entanglement of formation, given in Eq.~(\ref{diff}), against the purity of randomly created entangled states. It is apparent that the less the purity the larger the difference between $\mathcal{E}_F(\boldsymbol{\sigma}^{\text{sf}})$ and $\mathcal{E}_F^{\pm}(\boldsymbol{\sigma}^{\text{sf}})$. We also observe that on average the upper bound is closer to the real value than the lower bound.}
  \label{fig2}
\end{figure}

As we mentioned before, other methods of reaching the actual value for EoF have also been derived, but the one given in Eq.~(\ref{eof2}) is significantly easier for numerical calculations. A specific algorithm written in \textit{Mathematica} has also been developed \cite{supplementalmaterial}, that numerically evaluates the exact value of EoF for an arbitrary two-mode Gaussian state written in its standard form and parametrized according to Sec.\ref{sec2} A.

It worths also comparing the upper and lower bound to the actual value of EoF in order to see how close they are. In Fig.~\ref{fig2}, we randomly generate a large number of entangled states, and for each one we calculate the percentile relative difference, given by
\begin{equation}
\delta^{\pm}:=\frac{|\mathcal{E}_F-\mathcal{E}_F^{\pm}|}{\mathcal{E}_F}\times 100 \% \,,
\label{diff}
\end{equation}
against the global purity $\mu$ of the corresponding state. As we clearly see for a random state the purity is inversely proportional to the relative difference between $\mathcal{E}_F(\boldsymbol{\sigma}^{\text{sf}})$ and $\mathcal{E}_F^{\pm}(\boldsymbol{\sigma}^{\text{sf}})$. It is also apparent that the upper bound is on average closer to the exact value than the lower bound. Thus, besides the cases mentioned above, the upper and lower bounds can also be faithfully used for analytical calculations of the EoF for states with high purities, e.g., $ 0.8 \leqslant \mu \leqslant 1$. 

\section{Conclusions}
\label{sec4}

In conclusion, we derived an upper bound for the entanglement of formation for two-mode Gaussian states, that comes as an extension to the lower bound that we had recently derived in Ref.~\cite{Tserkis.Ralph.PRA.17}. The two bounds become tight for a wide range of states but they can also be considered quite faithful for highly pure states. We introduced a new method for computing the actual value of the entanglement of formation for two-mode Gaussian states, based on an optimization process over a single parameter, and we also provided a code written in \textit{Mathematica} for the numerical estimation of the measure for arbitrary two-mode Gaussian states \cite{supplementalmaterial}.

\section{Acknowledgments}

The research is supported by the Australian Research Council (ARC) under the Centre of Excellence for Quantum Computation and Communication Technology (CE170100012).

\end{document}